\documentclass[10pt]{article}   
\usepackage[a4paper,margin=1.4in]{geometry}

\usepackage[dvipsnames]{xcolor}
\usepackage{booktabs}
\usepackage{multirow}

\usepackage{graphicx}
\usepackage{ctable}

\definecolor{refcolor}{rgb}{0,0,0.5}
\ifpdf
\PassOptionsToPackage{hyphens}{url}
\usepackage[%
  pdftitle={Automated age-related macular degeneration area estimation - first results},%
  pdfauthor={Rokas Peciulis, Mantas Lukosevicius, Algimantas Krisciukaitis, Robertas Petrolis, Dovilė Buteikienė},%
  pdfkeywords={Age-related Macular Degeneration, Eye fundus image, Convolutional neural network, Deep Learning, ResNet50, ResNet101, MobileNetV3, UNet},%
  pdfstartview=FitH,%
  bookmarks=true,%
  bookmarksopen=true,%
  breaklinks=true,%
  colorlinks=true,%
  linkcolor=refcolor,%
  anchorcolor=blue,%
  citecolor=refcolor,%
  filecolor=blue,%
  menucolor=blue,%
  urlcolor=refcolor,%
  pdfpagelabels]{hyperref} 
\else 
\usepackage[%
  breaklinks=true,%
  colorlinks=true,%
  linkcolor=blue,%
  anchorcolor=blue,%
  citecolor=blue,%
  filecolor=blue,%
  menucolor=blue,%
  urlcolor=blue]{hyperref} 
\fi

\title{Automated age-related macular degeneration area estimation -- first results}

\author{
Rokas Pe\v{c}iulis$^{1,2}$ \\ \texttt{rokas.peciulis@ktu.edu} 
\and
Mantas Luko\v{s}evi\v{c}ius$^1$\href{https://orcid.org/0000-0001-7963-285X}{\includegraphics[scale=0.5]{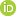}} \\ \texttt{mantas.lukosevicius@ktu.edu} 
\and
Algimantas Kriščiukaitis$^2$\href{https://orcid.org/0000-0003-4392-1937}{\includegraphics[scale=0.5]{orcid_16x16.png}} \\ \texttt{algimantas.krisciukaitis@lsmuni.lt} 
\and
Robertas Petrolis$^2$\href{https://orcid.org/0000-0003-3487-733X}{\includegraphics[scale=0.5]{orcid_16x16.png}} \\ \texttt{robertas.petrolis@lsmuni.lt} 
\and
Dovilė Buteikienė$^3$ \\ \texttt{dovile.buteikiene@lsmuni.lt}
}

\date{%
$^1$Faculty of Informatics, Kaunas University of Technology, Kaunas, Lithuania \\
$^2$Department of Physics, Mathematics and Biophysics, Lithuanian University of Health Sciences \\
$^3$Department of Ophthalmology, Mathematics and Biophysics, Lithuanian University of Health Sciences \\[2ex]%
        \today}

\begin{document}

\maketitle

\begin{abstract}
This work aims to research an automatic method for detecting Age-related Macular Degeneration (AMD) lesions in RGB eye fundus images. For this, we align invasively obtained eye fundus contrast images (the ``golden standard'' diagnostic) to the RGB ones and use them to hand-annotate the lesions. This is done using our custom-made tool. Using the data, we train and test five different convolutional neural networks: a custom one to classify healthy and AMD-affected eye fundi, and four well-known networks: ResNet50, ResNet101, MobileNetV3, and UNet to segment (localize) the AMD lesions in the affected eye fundus images. We achieve 93.55\,\% accuracy or 69.71\,\% Dice index as the preliminary best results in segmentation with MobileNetV3.
\end{abstract}

\section{Introduction}

The current prevalence of early Age-related Macular Degeneration (AMD) in Europe is 3.5\,\% in those aged 55-59 years and 17.6\,\% in those aged $>$85 years. It has decreased over the last 20 years but remains a significant public health problem \cite{colijn2017prevalence}. As it cannot be completely cured, early diagnosis and treatment can stop the progression and prolong the quality of the patient’s vision.

It is important to diagnose the AMD in the early stage when the damage to the retina is not very prominent. The evaluation of the area and anatomical localization of the lesion zone in AMD cases can help to control the treatment process.

Optic Coherence Tomography (OCT) imaging is used for visualization of AMD caused pathological neurovascularization zone, but the method is expensive and time-consuming \cite{suzuki2016microvascular}. AMD lesion zone detection and evaluation ``golden standard'' is fluorescein angiography -- a diagnostic method where sodium fluorescein dye is injected intravenously for pathological neovascularization visualization. However, this injection can have adverse effects, is costly, and invasive to the patient \cite{kwan2006fluorescein}. According to \cite{liang2010towards, yonekawa2015age} and our preliminary research results experienced ophthalmologists can detect certain visually noticeable features of RGB images related to AMD caused damage. However, there are no single or several distinct features to be used for all cases or their specificity is too low.

Superposition of registered fluorescence eye fundus images of AMD patients with RGB eye fundus images of the same patients can produce a training set for deep learning neural networks. The network eventually can be trained to detect and localize the lesion zone in RGB images.

There are successful examples of classification between AMD disease-affected and healthy eye fundi using OCT images \cite{lee2017deep} and attempts to automate the diagnosis of glaucoma using RGB eye fundus images \cite{raghavendra2018deep}.

This work aimed to assess the possibility of AMD lesion zone detection and evaluation in non-invasively registered RGB fundus images by using deep learning neural network architectures.

The objectives of our work are:
\begin{enumerate}
    \item Prepare the data:
    \begin{enumerate}
        \item Collect the AMD affected fundus images (RGB and contrast photos);
        \item Create the data preparation tool;
        \item Align RGB and contrast images;
        \item Annotate the lesion zone on the contrast images;
        \item Prepare comparable images without AMD.
    \end{enumerate}
    \item Investigate automated algorithms:
    \begin{enumerate}
        \item Train algorithms that can classify between images with and without AMD;
        \item Train algorithms that can segment the lesion area from the RGB image;
        \item Investigate performance of the algorithms and their applicability in real life.
    \end{enumerate}
\end{enumerate}

We present our materials and methods in Section \ref{section:methods}, introduce our dataset and its preparation in Section \ref{dataset}, present our algorithm investigations in Section \ref{methods}, the obtained preliminary results in Section \ref{results}, and conclude with a discussion in Section \ref{discussion}.

\section{Materials and Methods}\label{section:methods}

We have gathered a collection of RGB and gray-scale contrast images of eye fundi. We discuss the details of this data and its preparation in Section \ref{dataset}.

\subsection{Dataset}\label{dataset}

44 sets of age-related macular degeneration examination images were collected from the Lithuanian University of Health Sciences Department of Ophthalmology. All these sets contain AMD-affected cases of varying degrees.

Any personal information was removed and made sure that no identifiable details were left in file names or possible additional text files. The pictures from the same patient were grouped, making preliminary sets that contain a single RGB image and several contrast-enhanced images for damaged area detection. 

The set of images containing AMD had to be prepared for training. The RGB and contrast-enhanced images that were taken are misaligned by position, rotation, and scaling. In our case, we assumed that the tilt of the camera was not changed so only 3 latter parameters were used to match the images. The main reference points in matching images were the most visible vascular structures. To achieve a more convenient and effective image matching a specialized tool was created which is discussed in Section \ref{image-match-tool}.

To investigate automatic classification between healthy and AMD-affected images, we additionally took 15 healthy fundus images from Pattern Recognition Lab online database \cite{budai2013robust}.

\subsubsection{Segmentation Dataset Preparation Tool}\label{image-match-tool}

\begin{figure}
  \centering
  \includegraphics[width=\linewidth]{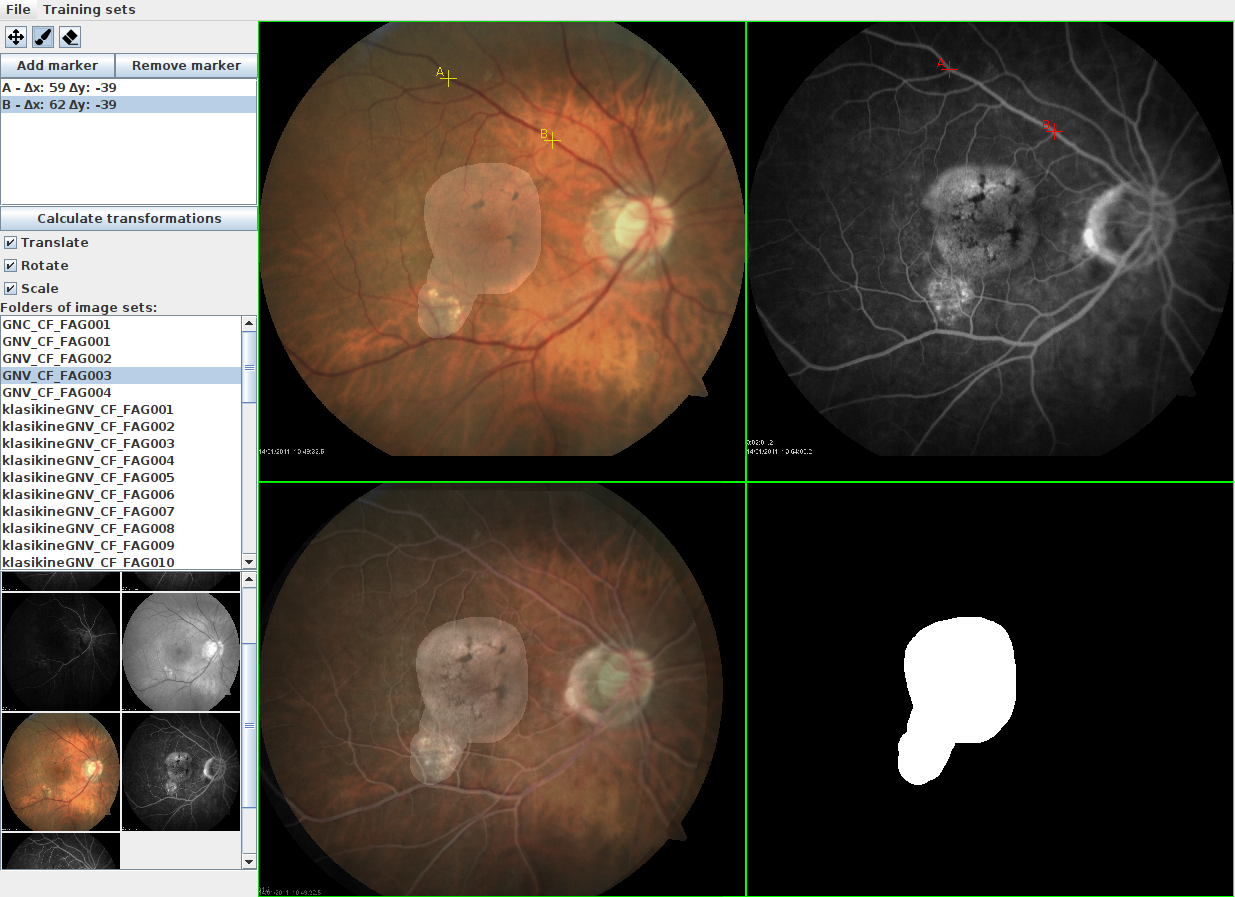}
  \caption{A screenshot of image annotation and processing tool created for our training dataset preparation. The main window is divided into 4 sections. The top-left section is the RGB image and the top-right section shows the contrast-enhanced image. The bottom-left section represents RGB and contrast-enhanced images matched and mask layer overlay. Finally, the bottom-right section shows the mask layer only. The RGB and mask images form a single training dataset. }
  \label{fig:image-tool}
\end{figure}

Before training data preparation, a convenient tool was created (Figure \ref{fig:image-tool}) to efficiently annotate and process raw data into training datasets for machine learning. The tool has the following functions:
\begin{enumerate}
    \item Preview raw images;
    \item Create a training set;
    \item Annotate the training set;
    \item Upload saved training sets to the online database;
    \item Download other training sets from the online database for preview and editing.
\end{enumerate}
The main task of this tool is to produce a matched and uniformly sized collection of training data sets for later use in neural network model training. The training data must have RGB images where all pixels are marked with ``ground truth'' information as being in a lesion or unaffected zone.

\subsubsection{Segmentation Dataset Preparation Process}

To create a dataset first the RGB and contrast-enhanced image (contrast image) are loaded into our dataset preparation tool. The tool lets a user add reference points in both images and applies contrast image transformation according to the reference points. If the resulting image is matched the user can continue painting the mask of the lesion area according to the matched contrast image. After this step, the tool produces an annotated and prepared training set which can be uploaded for the machine learning algorithm to train.

\subsubsection{Segmentation Training Dataset Structure}

\begin{figure}
  \centering
  \includegraphics[width=\linewidth]{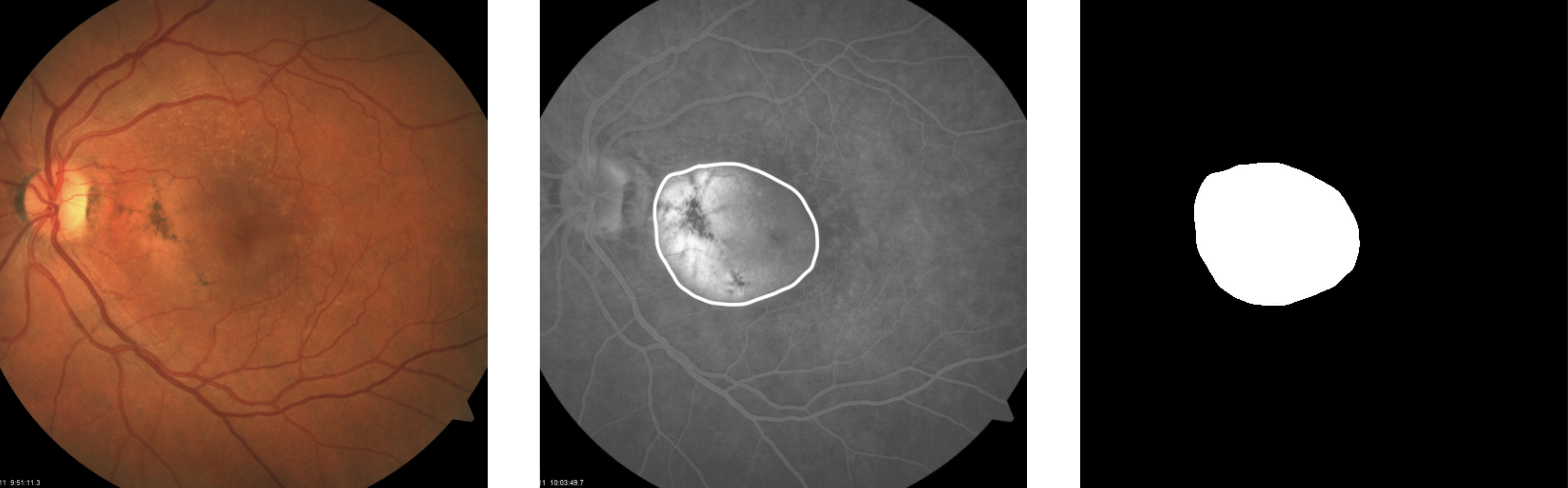}
  \caption{The training set image example. An RGB image is an input data representation is displayed on the left. In the middle there is a contrast enhanced image where the lesion zone is marked. On the right side the mask image is displayed which is the desired result for our lesion zone detection algorithm.}
  \label{fig:training-set}
\end{figure}

A single training dataset featured in Figure \ref{fig:training-set} contains an RGB image, a contrast-enhanced image, and a mask image which is in grayscale. The RGB image is the input data for algorithm training. The mask image corresponds to the lesion area of the retina and also is the desired output result after algorithm training. All dataset images are in preferred PNG format and 512x512 pixels resolution.

\subsubsection{Classification Dataset Preparation}\label{class_data}

For the classification task, we needed to have both healthy and AMD-affected images. Since the two classes came from different sources, we took some steps to make the images more similar.

We scaled, centered, and cropped the images to make them the same resolution. One obvious difference that remained, was the different dominating tints of the two datasets. To eliminate this, we included the experiments with color histogram equalization of the images as the pre-processing step. 

\subsection{Detection Algorithms}\label{methods}

In our evaluation process, we investigate two different algorithms for classification and segmentation respectively. The classification algorithm detects whether AMD is present in the given RGB image at all. The segmentation algorithm detects the location of the AMD lesion zones by producing a gray-scale output image highlighting them. The segmentation part of the process is performed only if the classifier detects AMD.

\begin{figure}
    \centering
    \includegraphics[width=0.3\textwidth]{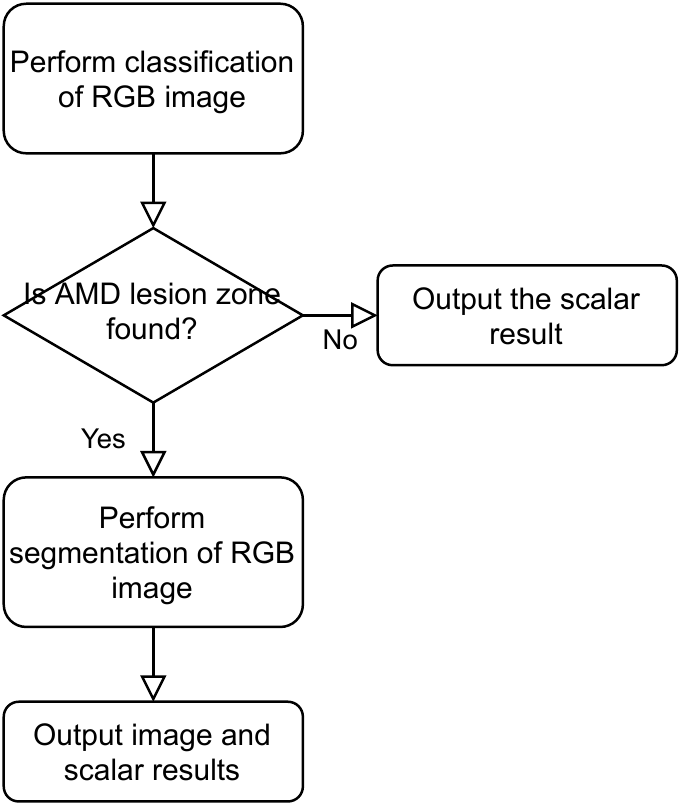}
    \caption{The flowchart of our evaluation process. At first, a classification algorithm evaluates the RGB image and returns a scalar value between 0 and 1 to evaluate a lesion zone is found (1 is positive, 0 negative). Our threshold was set to 0.5 to convert to Boolean values. If no lesion is found, only the scalar result of the calculated value is returned. If the lesion zone is detected, the segmentation of the RGB image is performed and the final result is both scalar and image results. The image result can then be overlapped with the RGB image to visualize the lesion zone more clearly. }
    \label{fig:evaluation_flowchart}
\end{figure}

The evaluation process flowchart is displayed in Figure \ref{fig:evaluation_flowchart}.
Each algorithm is described in more detail in Sections \ref{section:classification} and  \ref{section:segmentation} respectively.

\subsubsection{Degeneration Classification}\label{section:classification}

\begin{figure}
  \centering
  \includegraphics[width=0.65\textwidth]{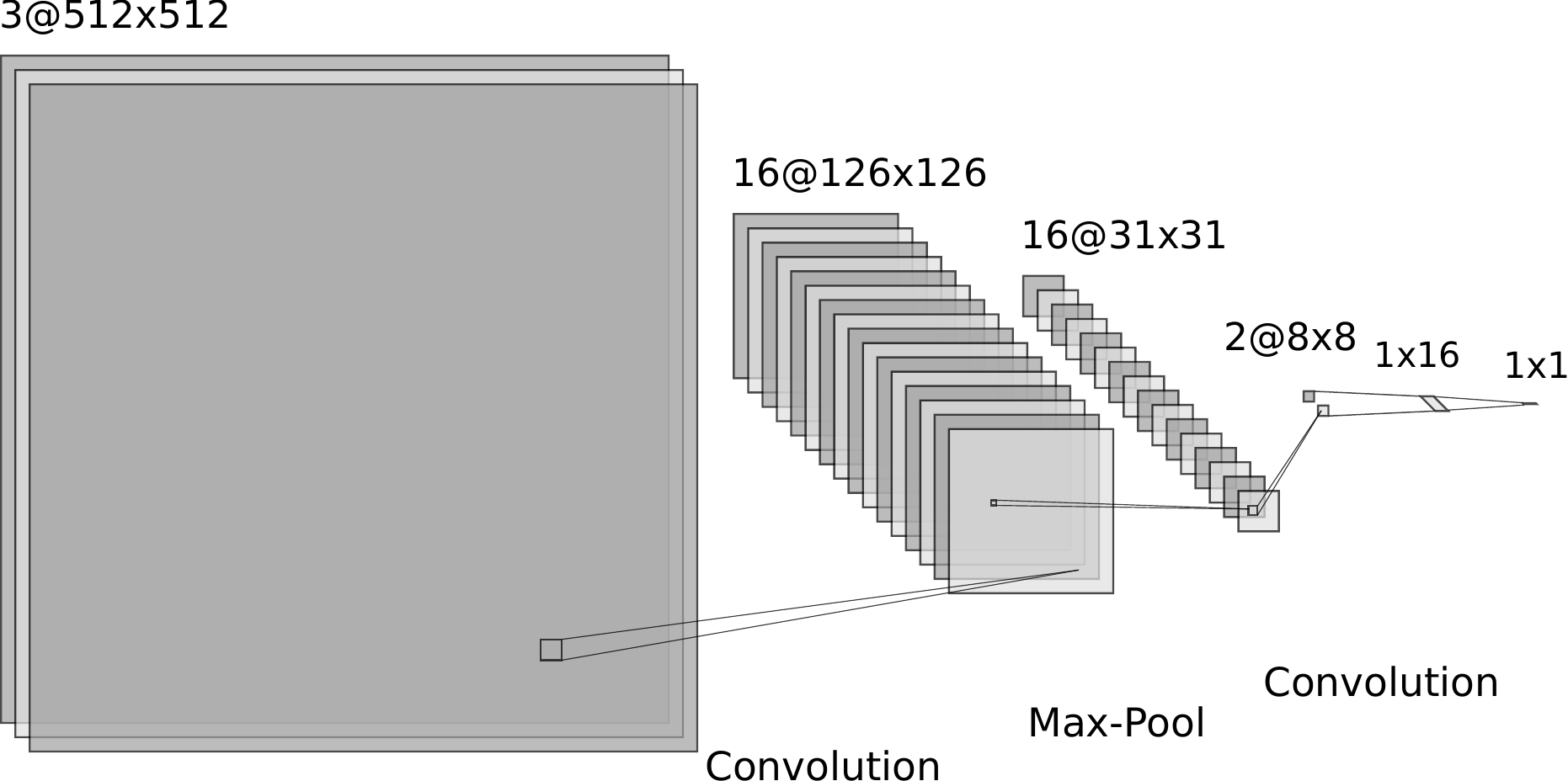}
  \caption{Architecture of the classification neural network.}
  \label{fig:nn_classification}
\end{figure}

For degeneration classification, a custom convolutional neural network was created which consists of 5 layers: 2 convolutions, 1 max-pooling, and 2 fully connected layers. The network takes the RGB image as the input and has a single output estimating the probability of degeneration. A detailed structure of classification neural network is presented in Figure \ref{fig:nn_classification}.

\subsubsection{Degeneration Segmentation}\label{section:segmentation}

For the degeneration segmentation algorithm, four well-known neural network architectures were tried: ResNet50, ResNet101 \cite{canziani2016analysis}, MobileNetV3 \cite{howard2019searching}, and UNet \cite{buda2019association}. The resolution for ResNet50, ResNet101, and MobileNetV3 neural networks is 384x384 pixels. For those networks the input data images were scaled at run-time and the output result was scaled back to 512x512 pixels. 

The models were trained from scratch (random weight initialization). The training configurations of the segmentation neural networks are presented in Table~\ref{tab:segmentation_configurations}.

\begin{table}[h]
  \centering
    \caption{Semantic segmentation neural network configurations for training.}
    \label{tab:segmentation_configurations}
    \begin{tabular}{lrrrr}
    \toprule
    Network model & Trainable parameters & Epochs & Batch size & Learning rate  \\
    \midrule
    ResNet50    & 39.6M  & 150 & 4 & 0.0001 \\
    ResNet101   & 58.6M  & 120 & 2 & 0.0001 \\
    MobileNetV3 & 32.2M & 150 & 4 & 0.0001 \\
    UNet        & 7.7M & 150 & 4 & 0.0001 \\
    \bottomrule
    \end{tabular}
\end{table}

\subsubsection{Measuring Detection Quality}

The confusion matrix and its derived measures were used to assess the detection quality of both the classification and localization algorithms: specificity, sensitivity, and accuracy measures were calculated. For lesion semantic segmentation Sørensen–Dice (Dice) coefficient, a.k.a. F1 score, was also calculated.

Before computing these measurements, we binarized the gray-scale output images to only black and white pixels. We used four different thresholds of 0.01, 0.05, 0.1, and 0.5 to compare which threshold worked best for each semantic segmentation neural network.

\section{Results}\label{results}

The results of our classification algorithm are presented in Table~\ref{tab:classification_results}. We can see that with histogram-normalized images, 99.93\,\% accuracy in the classification of images containing AMD lesions vs. healthy control can be achieved. 

\begin{table}[h]
 \centering
  \caption{Calculated performance of our classification algorithm in percent.}
  \label{tab:classification_results}
  \begin{tabular}{lrrr}
    \toprule
    \multirow{2.5}{*}{Parameter} & \multicolumn{2}{c} {AMD lesion detection} \\
    \cmidrule(rl){2-3}
     & Raw & Hist. equal. \\
    \midrule
    Sensitivity &   99.67   &   99.87   \\
    Specificity &   99.95   &   99.99   \\
    Accuracy    &   99.79   &   \textbf{99.93}   \\
    \bottomrule
  \end{tabular}
\end{table}

The results of our semantic segmentation models are presented in Table \ref{tab:segmentation_results}.
We can see that the best Dice result was achieved with the lowest threshold when binarizing the output images in all neural networks except for MobileNetV3. The best Dice result of 69.71\,\% was achieved with it and a 0.05 threshold. 

\begin{table}[h]
 \centering
  \caption{Calculated performance of used semantic segmentation algorithms in percent.}
  \label{tab:segmentation_results}
  \begin{tabular}{lrrrrrr}
  \toprule
  Network model & Threshold & Sensitivity & Specificity & Accuracy & Dice \\
  \midrule
  ResNet50 & 0.01 & 66.75 & 93.93 & 90.52 & 61.41 \\
           & 0.05 & 57.13 & 96.51 & 91.71 & 59.78 \\
           & 0.1  & 53.23 & 97.12 & 91.80 & 57.92 \\
           & 0.5  & 44.96 & 98.18 & 91.74 & 52.67 \\
  \midrule
  ResNet101 & 0.01 & 51.40 & 97.15 & 91.54 & 55.42 \\
            & 0.05 & 47.20 & 97.82 & 91.64 & 52.95 \\
            & 0.1  & 45.52 & 98.08 & 91.67 & 51.95 \\
            & 0.5  & 40.97 & 98.66 & 91.66 & 48.86 \\
  \midrule
  MobileNetV3 & 0.01 & 90.75 & 67.68 & 70.38 & 41.61 \\
              & 0.05 & 69.06 & 96.94 & 93.46 & \textbf{69.71} \\
              & 0.1  & 62.99 & 97.86 & \textbf{93.55} & 67.76 \\
              & 0.5  & 48.81 & 98.89 & 92.83 & 59.66 \\
  \midrule
  UNet & 0.01 & 41.33 & 96.94 & 90.60 & 49.63 \\
       & 0.05 & 39.71 & 97.20 & 90.65 & 48.66 \\
       & 0.1  & 39.00 & 97.30 & 90.67 & 48.19 \\
       & 0.5  & 37.32 & 97.60 & 90.80 & 47.19 \\
  \bottomrule
  \end{tabular}
\end{table}

\section{Discussion}\label{discussion}

The very high classification accuracy is most likely since the images of the AMD-affected and the healthy eye fundi are taken from different sources - obtained from different populations using different equipment.

We tried to make them similar, but this is probably impossible to do completely. Firstly, we cropped the images to mitigate different framing. But the color differences between the healthy and affected images were still visible by the naked eye. To eliminate these differences as a likely telltale sign for the classification, we equalized the color histograms of all the images, as mentioned in Section \ref{class_data}. However, this produced an even better classification (Table~\ref{tab:classification_results}).

In addition to equipment-based differences such as different color grading, lighting, position, and possibly tilt, the biggest difference might be in the subject population. While this is not indicated in the control dataset, it appears that the subjects were young, and the AMD-affected images are taken from old people.

To overcome this, we would need to collect and prepare more training data including healthy eye fundus images collected under the same conditions as the affected ones. But this is difficult to achieve because the procedure of taking eye fundus images requires pupil dilation medicine and is not normally performed on healthy subjects.

Overall, the results of the classification part of our algorithm are not so important to ophthalmologists, but are useful to our lesion zone segmentation algorithm, improving its performance. We plan to further concentrate our effort on the segmentation part.

\begin{figure}[h]
  \centering
  \includegraphics[width=\linewidth]{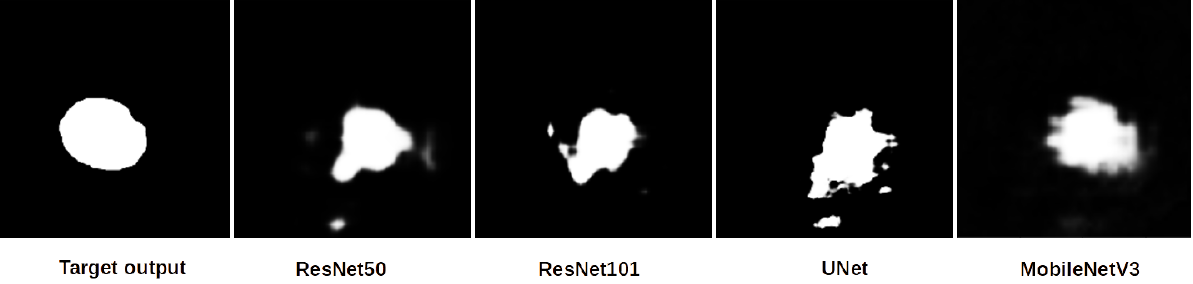}
  \caption{An example of the output images of the trained segmentation networks.}
  \label{fig:segmentation-images}
\end{figure}

While observing typical outputs produced by our tested segmentation neural networks in Figure~\ref{fig:segmentation-images}, we can see that the UNet model produces very uneven and disconnected annotations. They differ greatly from the target output and result in the worst performance. The other models produce more blurry-edged annotations but they are more homogeneous. This is most true for MobileNetV3, which gives the best performance, followed by ResNet50 and then ResNet101.

The result of the segmentation algorithm is important for periodic checkups of a patient who has been diagnosed with AMD to objectively observe the progress of the disease. Such a method could help to recognize the AMD progress without invasive contrast-enhanced angiography or costly OCT imaging. The detected AMD lesion zone can be further used to automatically compute the lesion area ratio to the healthy retina and the previous scans of the patient, to give a quantitative estimate of the progress of the disease.

\bibliographystyle{unsrt}     
\bibliography{sample-ceur}

\end{document}